\documentclass[runningheads]{llncs}

\usepackage[T1]{fontenc}
\usepackage[utf8]{inputenc}  
\usepackage{pgf} 
\usepackage{graphicx}
\usepackage{array}
\usepackage{ragged2e}
\usepackage{physics}
\usepackage{amsmath}
\usepackage{amssymb}

\usepackage{soul}
\usepackage{todonotes}

\DeclareUnicodeCharacter{03B2}{\ensuremath{\beta}}  

\begin{document}
\title{Auto Quantum Machine Learning for Multisource Classification}

\author{Tomasz Rybotycki\inst{1, 2, 3}\orcidID{0000-0003-2493-0459} \and
	Sebastian Dziura\inst{1}\orcidID{0009-0008-6786-511X} \and
	Piotr Gawron\inst{1, 2}\orcidID{0000-0001-7476-9160}}
\authorrunning{T. Rybotycki, S. Dziura, P. Gawron}
%
\institute{
	Center of Excellence in Artificial Intelligence, AGH University, al.~Mickiewicza 30, 30-059 Cracow, Poland
	\and
	Nicolaus Copernicus Astronomical Center, Polish Academy of Sciences, ul. Bartycka 18, 00-716 Warsaw, Poland
	\and
	Systems Research Institute, Polish Academy of Sciences, ul.~Newelska 6, 01-447 Warsaw, Poland
}
\maketitle              
\begin{abstract}
With fault-tolerant quantum computing on the horizon, there is growing interest in applying quantum computational methods to data-intensive scientific fields like remote sensing. Quantum machine learning (QML) has already demonstrated potential for such demanding tasks.
One area of particular focus is quantum data fusion—a complex data analysis problem that has attracted significant recent attention.
In this work, we introduce an automated QML (AQML) approach for addressing data fusion challenges. We evaluate how AQML-generated quantum circuits perform compared to classical multilayer perceptrons (MLPs) and manually designed QML models when processing multisource inputs.
Furthermore, we apply our method to change detection using the multispectral ONERA dataset, achieving improved accuracy over previously reported QML-based change detection results.


\end{abstract}
%
%
\section{Introduction}
Multisource Data about an observed object can be gathered from multiple
complementary sensors. These data sensors have to be
aggregated and transformed into useful information to e.g.\ obtain richer and
more accurate decision outcomes. The field of multisource information fusion (MSIF) deals with this type of data processing problems.

In this work, we focus on
feature-level fusion, in which the fusion model operates on extracted feature
representations. MSIF has found widespread application in machine learning,
particularly in classification tasks, as surveyed
in~\cite{baltrusaitisMultimodalMachineLearning2019}. Representative applications
include remote sensing~\cite{liA3CLNNSpatialSpectral2022}, military
systems~\cite{sycaraIntegratedApproachHighlevel2009}, and human action
recognition~\cite{fortinoAdvancesMultisensorFusion2019}, among
others~\cite{liMultisourceInformationFusion2024}.

The issue mentioned above lies at the very core of the MSIF, a domain of
knowledge dedicated to finding efficient methods for multisource data aggregation. The field itself has been a
subject of scientific investigations for quite some time now, particularly in the context of remote sensing
\cite{DataFusion1986}, where it persists as a subject of studies \cite{liA3CLNNSpatialSpectral2022}. The research
on data fusion is extensive enough, and the scope of techniques it presents is so wide, that they could be
categorized into groups and further subgroups \cite{baltrusaitisMultimodalMachineLearning2019}. In this work, we
focus solely on one such subgroup: a model-agnostic early fusion, otherwise known as feature-based fusion. This
term refers to a family of algorithms in which the fusion occurs at the early stages of the ML pipeline. A subtype of
those algorithms is a data-level fusion, an approach in which the multisource data is aggregated prior to entering
the ML model \cite{dalla_mura_challenges_2015}.

Early fusion problems are often approached with machine learning models \cite{liMultisourceInformationFusion2024,baltrusaitisMultimodalMachineLearning2019}, which one may consider a very
natural approach. After all, ML models are trained on the data. If the data is represented better (with multiple
representations, coming from multiple sources), it is only natural that the results should improve. However, some of
the data fusion problems have been shown to be NP-hard \cite{DataFusionNPQAOA}. This means that classical techniques,
while still useful, might not be able to address the problems above certain sizes (unless we see some unlikely shift
in the complexity theory). This is exactly why alternative methods of computing, such as quantum computing and
quantum machine learning (QML), have recently been of interest for data fusion specialists.

Quantum machine learning (QML) leverages principles of quantum mechanics to process information in ways that differ
fundamentally from classical machine learning~\cite{biamonteQuantumMachineLearning2017a}. In this paradigm,
computations are done within quantum circuits, commonly with the use of quantum logic gates (see e.g.
\cite{sebastianelli2021circuitbasedhybridquantumneural}). In the mathematical model of quantum computation, those
gates can be expressed by unitary gates, making the whole circuit (up to the measurement) a unitary transformation
of the input. However simple the description may look, quantum mechanics introduces two computationally useful
phenomena --- superposition and entanglement --- that when used correctly can significantly speed up the
computations. That said, we can think of a QML model as a ML model that uses quantum computing. It is known that
such models can represent complex functions using comparatively shallow (i. e. with low layers count) circuits
or fewer parameters than their classical counterparts, suggesting potential advantages for learning tasks
involving structured or high-dimensional
data~\cite{havlicekSupervisedLearningQuantumenhanced2019,schuldQuantumMachineLearning2019,abbasPowerQuantumNeural2021,GawronPowerOfData}. Moreover,
theoretical work has identified learning scenarios in which quantum models offer data-dependent
advantages~\cite{huangPowerDataQuantum2021a}. Empirical tests on the multispectral satellite data seem to also
confirm this finding \cite{GawronPowerOfData}.


Recently, several works explored the application of QML methods to multisource or multi-modal data fusion.
Common approach is to, instead of using purely-quantum models, take advantage of the hybrid quantum-classical
models. Feature-level multi-modal quantum fusion presented in \cite{wuFeatureEntanglementbasedQuantum2026} is a
stellar example of that approach. The ML pipeline described therein, takes features extracted by separate
uni-modal classical neural networks and embeds them into a quantum fusion circuit. The features are then entangled
between multiple modalities on target qubits. Their model showed consistent stability and high accuracy for
high-dimensional data.

When it comes to QML models, one of the central problems is quantum architecture search (QAS) \cite{QAS}. It is a
quantum version of the neural architecture search (NAS) problem known in the ML community. Without the right
selection of the model architecture, the training might fail. While the problem case of both NAS and QAS, there are
several techniques that allow to address it. One of such techniques is automated (quantum) machine learning (AQML).
AQML are heuristic algorithms that try to propose the best models for a given data and tasks. There are several
platforms for AQML (see \cite{Rybotycki_2025} for brief overview) available for use and reports of the
state-of-the-art results achieved using at least one of them \cite{AQML_Exp}.

It this paper, we present an application of AQML to quantum feature fusion problem. In the following Section
\ref{sec:problem_formulation} we formally present the issues addressed in this manuscript: the feature fusion and
quantum architecture search. Then, we provide the detailed description of our multi-level approach in Section
\ref{sec:method}. Section \ref{sec:exp} contains the experiments overview --- along with the datasets description,
hyperparameter optimization (HPO) step and the results' analysis. We conclude the paper with final remarks and
insights for future work.

\section{Problem formulation and literature overview}
\label{sec:problem_formulation}

Performing the feature fusion using QML models is a complex, multi-step problem. While it most notably involves a QAS problem,
there are also multiple other decisions to be made. Complete pipelines are not necessarily restricted
to quantum models, after all. In this Section, we detail the feature fusion problem and further argument why
QML models are plausible candidates for providing satisfactory solutions in that regard. We also introduce a
specific case of data fusion --- change detection task --- that we considered in our experiments with the remote
sensing satellite data. Finally, we formally introduce QAS problem, and discuss its usual constraints that we won't
consider.

\subsection{Data fusion}

Data fusion has a goal-oriented definition.
\begin{center}
	\textit{Join information from multiple modalities (sources) to perform prediction \cite{baltrusaitisMultimodalMachineLearning2019}.}
\end{center}
In this work we study early fusion, which means the data will be aggregated directly after its extraction. Data
fusion is known to enhance the quality of MLP model predictions and is required for tasks such as change detection
\cite{QMLCaseStudy}. The simplest form of data fusion is concatenating the extracted feature vectors.

\subsection{Change detection}

Change detection can be defined as the following task.
\begin{center}
	\textit{Given a pair of corresponding pictures, depicting the same region and taken at different times, find the
		regions that differ between the pictures \cite{QMLCaseStudy}. }
\end{center}
Change detection can be considered a multisource data fusion problem, even when the picture was taken using
the same device. This is due to the fact that the temporal difference between the images implies (in general)
different atmospheric conditions, and notable spatial (angles, distances, ...) or geographical differences (new
buildings, ...). This, in turn, introduces heterogeneity to the pair of pictures. In the terms of data processing,
change detection can be seen as a semantic segmentation task \cite{Segmentation}, or even as binary classification
(change or no-change) \cite{QMLCaseStudy}. Change detection is a well studied problem, especially in remote
sensing~\cite{ChangeDetectionSurvey}.

\subsection{Quantum Architecture Search}

Quantum architecture search is a problem known by many names, and refers to automatic selection of quantum circuits
appropriate for given task \cite{QAS}. Formally speaking, one can formulate the QAS problem as follows.
\begin{center}
	\textit{Given the task and the search space, find within the search space a quantum circuits that best fits the task,
		according to predefined evaluation criteria.}
\end{center}
Most often, the search space is defined by the target device, on which the circuit is meant to run. This restricts
the building blocks of the circuit to a set of gates native to the device. It also limits the number of qubits
one can use and defines the effective maximal depth of the circuit. In the current noisy intermediate-scale
quantum era, device selection is vital to QAS. So much so that many of the proposed methods are hardware-aware
by nature \cite{QAS}, and in the context of QML a notion of device-oriented training has been mentioned
\cite{QMLCaseStudy}. The authors of \cite{QMLCaseStudy} also report how significant were the differences when the
model trained on a noiseless simulator was run on a real quantum device.

\section{Approach}
\label{sec:method}

In this work, we present two different approaches, sharing a common core in the form of AQML. We detailed both in
Subsection \ref{ssec:hqnn}. After all the models are introduced, we will briefly describe the AQML technique we used
and direct interested readers to appropriate resources. In particular, we focus on the adjustments that are new in
comparison to its reported version \cite{Rybotycki_2025}.

While the models we used differ slightly from each other, we would like to point out that their underlying
high-level logic remains the same and it is only the implementation aspect that differs. Each of our models
consists of two main parts: feature extractor and a classifier. The feature extractor is the initial part of
the model we used. It accepts the input and initially processes the data. For every model, feature extractor
we used is classical. Preprocessed features are then passed to the classifier. The output of this
model will be used to obtain the final decision about the input, depending on the experiment. This part of the
network may be classical (Subsection \ref{ssec:mlp}) or quantum and hybrid quantum-classical (Subsection
\ref{ssec:hqnn}).

\subsection{Multi-layer perceptron}
\label{ssec:mlp}

When defining the model for our initial quantum data fusion experiments, we had to make a decision regarding
which part of the model we should make quantum. Following logic similar to that in
\cite{wuFeatureEntanglementbasedQuantum2026}, we decided to use quantum layers for our classifier. Moreover,
considering the effectiveness of the network in \cite{wuFeatureEntanglementbasedQuantum2026}, we also
decided to separate the feature extractor part of the model into two distinct parts --- one for each input.

\begin{figure}
	\centering
	\hspace*{-0.07\paperwidth}\includegraphics[width=1.2\textwidth]{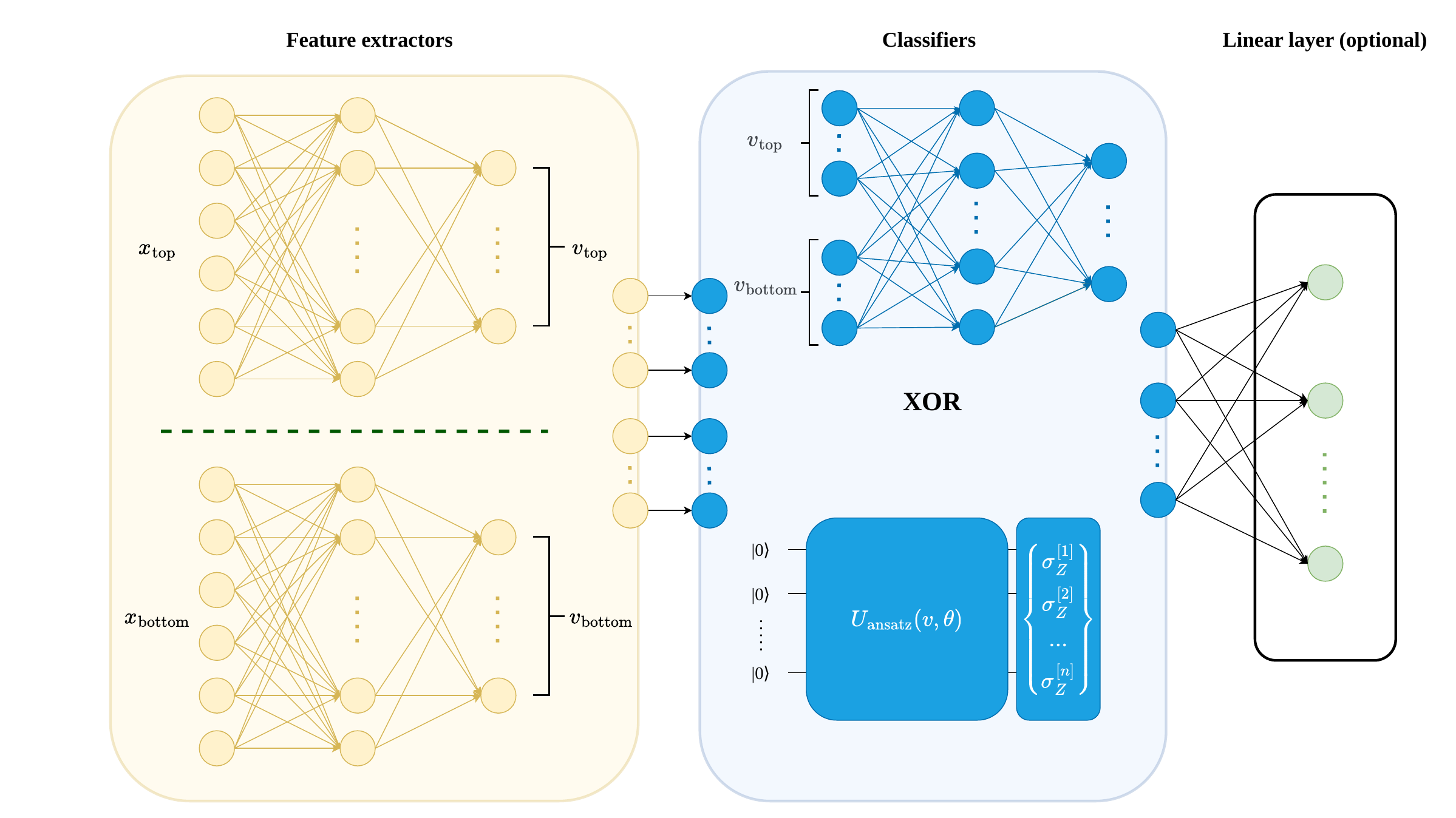}
	\caption{A schematic representation of all the models used in this work: \textbf{yellow} --- feature extractor, \textbf{blue} --- classifier (here the fusion takes place), \textbf{green} --- optional output logits for quantum
		classifiers. XOR on the schematic means that only one type of classifier is used at the time. Top classifier
		represents MLP approach, and the bottom one parametrized quantum circuit approach.}
	\label{fig:model_schamatic}
\end{figure}

The first model we implemented was a classical multi-layer perceptron (MLP) \cite{RussellNorvig2010}, which was meant to
serve as a baseline reference for our subsequent experiments. In this model both the extractors and the classifier were
a MLP. We present this a architecture in the Figure~\ref{fig:model_schamatic}.
\vspace*{-0.5cm}

In this architecture, the inputs from both sources --- $x_\mathrm{top}$ and $x_\mathrm{bottom}$ --- are passed to
their respective feature extractors and then processed, independently. After this processing, the output vectors of
both extractors are concatenated, such that $v = [v^1_\mathrm{top}, ..., v^n_\mathrm{top}, v^1_\mathrm{bottom}, ...,
	v^m_\mathrm{bottom}]$, assuming the lengths of $v_\mathrm{top}$ and $v_\mathrm{bottom}$ are $n$ and $m$ respectively.
The resultant vector $v$ is used as the input of the classifier network (here another MLP). For all the layers (beside
the output) we used $ ReLU(v) = \max(v, 0)$ activation function, which is a standard way to introduce non-linearity in
deep neural networks \cite{RussellNorvig2010}.

\subsection{Hybrid Quantum--Classical Neural Network}
\label{ssec:hqnn}

In our initial experiments with quantum machine learning, we used a hybrid quantum-classical machine learning
model. What the name entails is the fact that in the ML pipeline, both classical and quantum layers are used. A common
approach is to include a parameterized quantum circuit (PQC), with an arbitrary number of gate layers, as a single
quantum layer of the network. Such layer can then be a part of a model, or a model by itself.

Starting from the model schematic, as presented in the Figure \ref{fig:model_schamatic}, we kept the feature extractors
as in the MLP experiment (see Subsection \ref{ssec:mlp}). However, we used quantum classifier, instead of the classical
one. This way, our final architecture resembled the one in \cite{wuFeatureEntanglementbasedQuantum2026}. Moreover, this
way we could assess the classification capabilities on the found models, under the assumption that the data
preprocessing (i.e. the preprocessing algorithm) remained intact.

\begin{figure}
	\centering
	\includegraphics[width=0.6\textwidth]{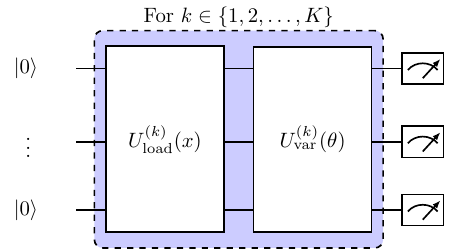}
	\caption{A general schematic of a parameterized quantum circuit. It consists of $K$ QML blocks, each containing a
		data (re-)uploading operation and a variational operation. Notice that both $U_{\mathrm{load}}$ and $U_{\mathrm{var}}$
		could, in principle, be an identity. On the schematic, $x$ denotes the input, and $\theta$ denotes the full set of PQC
		parameters. The PQC is concluded with a measurement.}
	\label{fig:AQML_PQC}
\end{figure}

In the construction of our quantum classifiers, we used two different approaches. First we tried manual selection of the
ansatz (which is another usual name for a PQC). In that case, the model we selected had no data re-uploading and
consisted of typically used quantum neural network (QNN) layers. In our other approach, we used AQML techniques to
find the right ansatz for the problem. In this scenario, the algorithm suggested the circuit built using several QML
blocks, which are pairs of data (re-)uploading and variational operations. The resultant PQC schematic is presented in
the Figure \ref{fig:AQML_PQC}.
For some experiments, we enhanced this architecture with an additional classical layer
that processed the output of the PQC and naturally led to a desired number of outputs.

The importance of data uploading layers cannot be overstated. First of all, at least one such layer is required of any
QML model, as raw classical data cannot be processed by a quantum circuit. It has to be encoded into a quantum state.
Second of all, data re-uploading is known --- both theoretically and empirically --- to increase the expressiveness of
quantum circuits \cite{schuld_effect_2021}.

\subsubsection{Change detection hybrid model}
\label{sssec:qml_onera}

To investigate the performance of our approach with the real data, we decided to revisit the experiments reported in
\cite{QMLCaseStudy}, only this time, with the use of AQML. The model we used for those experiments follows the
schematic presented in the Figure \ref{fig:model_schamatic} and is similar to the model described in previous Subsection
\ref{ssec:hqnn}. The only difference here is that we used a single feature extractor instead of two, to which we uploaded
the concatenated data from both sources. We decided not to separate the feature extraction, so that our experiment
differs from \cite{QMLCaseStudy} only in the application of AQML. In this model, the final classical layer was also
an optional addition. To reduce the initial dimensionality of the input (with 13-band pixels it would require too many
qubits for efficient simulation) PCA was used, leaving the input from each source described by four features.

\subsection{Automated Quantum Machine Learning}

In our experiments, we used the \texttt{aqmlator} library and thus the hyperparameter optimization (HPO)-based AQML
approach reported in \cite{Rybotycki_2025}. The idea behind this algorithm is to treat the ansatz architecture as an
hyperparameter and apply classical HPO techniques to find it. However, instead of using quantum gates as the basic
building blocks of the PQC, the authors proposed to use well-known and commonly used quantum layers, and thus limiting
the search-space of the optimization algorithm significantly.

The latest version of \texttt{aqmlator} (2.0.0a1, available only through the sources), aside from technical
improvements, offers more flexible ansatze finding classes, proposes data re-uploading layers, and uses a wider range
of variational layers during the ansatz search. In particular, it implements the \texttt{SimplifiedTwoDesign} and the
\texttt{BellmanLayer}, thus making it possible for \texttt{aqmlator} to find the models used in the previous ONERA QML
change detection experiments \cite{QMLCaseStudy}.

\section{Experiments}
\label{sec:exp}

In the following sections, we present the overview of the quantum data fusion experiments we've done during this study.
We begin with the description of our initial experiments using synthetic data created from the well-known MNIST dataset
\cite{lecunMNISTHandwrittenDigits2025}. With the successful application of QML and AQML to the synthetic data, we shift
our attention to the real-world ONERA dataset \cite{onera} and revisit experiments from \cite{QMLCaseStudy}, but this
time using AQML techniques to select the quantum classifier. The code and the data from our MNIST experiments are
available through Zenodo \cite{dziura_2026_18717347}. Our ONERA AQML experiments are in the original repository of
\cite{QMLCaseStudy}.

\subsection{MNIST}

\subsubsection{Dataset}
The original MNIST dataset consists of 70,000 grayscale images of handwritten digits (0--9), each of size $28
	\times 28$ pixels. We obtained the dataset from Kaggle \cite{lecunMNISTHandwrittenDigits2025}, and used the
predefined training--validation split as-is.

In the preliminary research, we reduced the number of output classes from 10 to 3. The idea here was to reduce the
complexity of the task, so that we could ensure reasonable (i. e. simulable) sizes of our quantum classifiers. We
also wanted not to overcomplicate our preliminary examples. We arbitrarily selected:
\begin{itemize}
	\item digit ``5'' (\(5421\) training samples and \(892\) validation samples),
	\item digit ``6'' (\(5918\) training samples and \(958\) validation samples),
	\item digit ``7'' (\(6265\) training samples and \(1028\) validation samples).
\end{itemize}

\begin{figure}
	\centering
	\includegraphics[width=\textwidth]{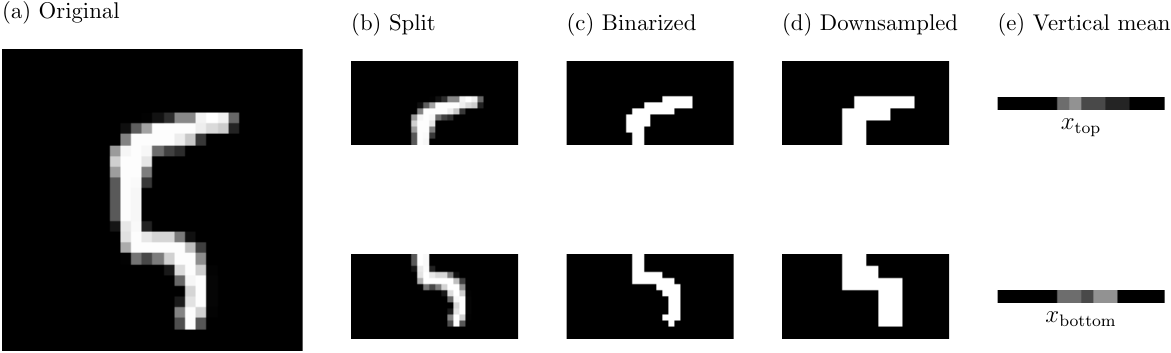}
	\caption{Overview of the synthetic multisource preprocessing pipeline.
		(a) Original input image.
		(b) Horizontal splitting into top and bottom halves.
		(c) Binarization of each half.
		(d) Spatial downsampling.
		(e) Computation of column-wise mean pixel intensities, resulting in vector representations
		that constitute the synthetic multisource inputs $x_{\mathrm{top}}$ and $x_{\mathrm{bottom}}$.
	}
	\label{fig:dataPrep}
\end{figure}

Each image was preprocessed to construct synthetic multisource inputs. Images were split horizontally into top and
bottom halves, which were treated as complementary views. Each view was then binarized, spatially downsampled, and
transformed into a 14-dimensional vector by computing the mean pixel intensity along the columns. These vectors
correspond to the synthetic multisource inputs $x_{\mathrm{top}}$ and $x_{\mathrm{bottom}}$, each described with
14 features. We graphically present this data preparation scheme in the Figure \ref{fig:dataPrep}.

\subsubsection{Experimental setup}

The $x_{\mathrm{top}}$ and $x_{\mathrm{bottom}}$ are used directly as the input to the respective feature
extractors. Each feature extractor had, therefore, 14 input neurons. The number of neurons in the hidden and output
layers of the feature extractors were determined by the HPO experiments detailed below. In those experiments,
feature extractors had only one hidden layer.

Similarly for the number of a single hidden layer neurons in the classical classifier and number of QML
blocks in the quantum classifier. Classifiers' number of inputs was determined by the length of concatenated output
of the feature extractors $v$, and output size was determined by the number of classes, which was fixed to three.
The reason for the latter was the use of standard one-hot encoding \cite{hancock_survey_2020}, common in
classification tasks.

The training goes as follows. The vector inputs is subsequently passed through learnable feature extraction
functions $f(\cdot,\theta_1)$ and $f(\cdot,\theta_2)$, producing feature vectors $v_{\mathrm{top}}$ and
$v_{\mathrm{bottom}}$ of equal dimensionality. These features are fused and classified by a function
$h(\cdot,\theta_3)$ --- realized by the classifier --- yielding the final classification decision $d$. The
resulting formulation is defined as:
\begin{align*}
	v_{\mathrm{top}}    = f(x_{\mathrm{top}}, \theta_1); \quad 
	v_{\mathrm{bottom}}  = f(x_{\mathrm{bottom}}, \theta_2); \quad
	d                    = h(v_{\mathrm{top}}, v_{\mathrm{bottom}}, \theta_3).
\end{align*}
All learnable parameters $\theta_1$, $\theta_2$, and $\theta_3$ are optimized jointly during the training. The
training is performed using the cross-entropy loss, and predictions are obtained by selecting the output
logit (or qubit) with the maximum output value. In the case of PQC, we used measured expectation value of the
Pauli $Z$ operator for each qubit. We trained all the models using 5-fold cross-validation, to better assess the
average performance of the investigated models. The optimizer we used was standard ADAM optimizer \cite{ADAM}.

\subsubsection{Metrics}
Although we gathered information about accuracy, precision, recall and f1-scores of the model, in their different
variants implemented in \texttt{scikit-learn} version 1.8, the scores we ultimately used for models evaluation were
accuracy and averaged f1$_{\mathrm{macro}}$ score. Both are standard metrics used for measuring classification
performance, and f1-score aggregates the information given by recall and precision (by definition). This set of
metrics allowed us to assess the general performance of the evaluated models.

\subsubsection{Hyperparameter Optimization}
\label{ssec:hpo}

We performed hyperparameter optimization (HPO) to select the optimal architecture for the models we tested in the
experiment. To that aim, we used \texttt{optuna} hyperparameter optimizer \cite{optuna}, a \texttt{Python} package
for HPO, used in both ML and QML. We present the HPO setup and the results for the classical and quantum experiments in
the Tables~\ref{tab:baseline_hpo} and \ref{tab:pqc_hpo}. Those results were obtained without early stopping, and
using random seed equal to 42. The number of trials varied, depending on the model, and was equal to 100, 100, 50
and 35 for MLP, $PQC_{\mathrm{manual}}$, $PQC_{\mathrm{AQML}^{\mathrm{Solo}}}$ and
$PQC_{\mathrm{AQML}^{\mathrm{Linear}}}$, respectively. The numbers differ due to limited compuational
resources.

\begin{table}[t]
	\centering
	\caption{Experiment setup for baseline hyperparameters optimization.}
	\label{tab:baseline_hpo}
	\begin{tabular}{p{6cm}p{2cm}p{1.5cm}l}
		\hline
		\textbf{Parameter}                     & \textbf{Range}        & \textbf{Type} & \textbf{Best value} \\
		\hline
		Learning rate                          & $[10^{-3},\,10^{-2}]$ & linear        & $10^{-3}$           \\
		Batch size                             & $[32, 128]$           & discrete      & $83$                \\
		Hidden layer size (feature extractors) & $[64, 256]$           & discrete      & $90$                \\
		Output layer size (feature extractors) & $[64, 256]$           & discrete      & $196$               \\
		Hidden layer size (classifier)         & $[64, 256]$           & discrete      & $95$                \\
		\hline
	\end{tabular}
\end{table}

\begin{table}[t]
	\centering
	\caption{Experiment setup for PQC hyperparameters optimization. For those experiments the learning rate was
		fixed to $lr = 10^{-3}$. PQC$^{\mathrm{Solo}}_{\mathrm{AQML}}$ and PQC$^{\mathrm{Linear}}_{\mathrm{AQML}}$ denote
		PQC models without and with final classical linear layer respectively. We present the architecture of the model
		in the parenthesis, next to the model's name.}
	\label{tab:pqc_hpo}
	\begin{tabular}{p{6cm}p{2cm}p{1.5cm}l}
		\hline
		\textbf{Parameter}                                 & \textbf{Range} & \textbf{Type} & \textbf{Best value}                                           \\
		\hline
		\multicolumn{4}{c}{PQC$_{\mathrm{Manual}}$ (\texttt{AngleX} $ \to 3 \times $ \texttt{BEL} )}                                                        \\
		\hline
		Batch size                                         & $[32, 128]$    & discrete      & $32$                                                          \\
		Hidden layer size (feature extractors)             & $[64, 256]$    & discrete      & $128$                                                         \\
		Output layer size (feature extractors)             & $[4, 12]$      & discrete      & $6$                                                           \\
		\texttt{BasicEntangler} layers number (classifier) & $[1, 5]$       & discrete      & $3$                                                           \\
		\hline
		\multicolumn{4}{c}{PQC$^{\mathrm{Solo}}_{\mathrm{AQML}}$ (\texttt{Amplitude} $\to$ \texttt{SEL} $\to$ \texttt{AngleY} $\to 2 \times$ \texttt{SEL})} \\
		\hline
		Batch size                                         & $[32, 128]$    & discrete      & 76                                                            \\
		Hidden layer size (feature extractors)             & $[64, 256]$    & discrete      & 163                                                           \\
		Output layer size (feature extractors)             & $[4, 12]$      & discrete      & 8                                                             \\
		QML Blocks number (classifier)                     & $[1, 5]$       & discrete      & 3                                                             \\
		\hline
		\multicolumn{4}{c}{PQC$^{\mathrm{Linear}}_{\mathrm{AQML}}$
			(\texttt{Amplitude} $\to$ \texttt{BEL} $\to$ \texttt{AngleX} $\to$ \texttt{BEL} $\to$ \texttt{AngleZ}
		$\to$ \texttt{BEL} $\to$ \texttt{SEL} $\to$ \texttt{BEL} )}                                                                                         \\
		\hline
		Batch size                                         & $[32, 128]$    & discrete      & 121                                                           \\
		Hidden layer size (feature extractors)             & $[64, 256]$    & discrete      & 97                                                            \\
		Output layer size (feature extractors)             & $[4, 12]$      & discrete      & 8                                                             \\
		QML Blocks number (classifier)                     & $[1, 5]$       & discrete      & 5                                                             \\
		\hline
	\end{tabular}
\end{table}

\subsubsection{Results}

In the Table~\ref{tab:mnist_results} we present the values of metrics obtained with the best classical, manual PQC and
AQML-found PQC models. The results clearly show that the models found with AQML outperformed manually selected PQCs
for our synthetic multisource MNIST classification task. The accuracy they obtain is on par with the classical MLP,
but with over 10 times less parameters. Over 100 times less, if we consider only classifiers.

In the Table~\ref{tab:mnist_results} one can also see that while PQCs with additional linear layer performed better,
the difference between the best models reported isn't significant (is within a single standard deviation). We have
noticed, however, that the models with linear layer seem to behave more consistently during the training. We started
to investigate that phenomenon, and found that indeed accuracy averaged over all models from given groups was:
\begin{equation}
	\mathrm{accuracy}^{\mathrm{avg}}_{\mathrm{PQC}^{\mathrm{Solo}}_{AQML}}
	= 0.837 < 0.889 =
	\mathrm{accuracy}^{\mathrm{avg}}_{\mathrm{PQC}^{\mathrm{Linear}}_{AQML}},
\end{equation}
which seems to confirm our initial insights. It's also consistent with another empirical observation:
\begin{equation}
	\mathrm{accuracy}^{\mathrm{min}}_{\mathrm{PQC}^{\mathrm{Solo}}_{AQML}}
	= 0.367 < 0.631 =
	\mathrm{accuracy}^{\mathrm{min}}_{\mathrm{PQC}^{\mathrm{Linear}}_{AQML}}.
\end{equation}
This fact was also observed for the other metrics. We plan to continue to investigate this issue further as a part
of a separate study.

\begin{table}[t]
	\centering
	\caption{Average accuracy and f1$_{\mathrm{macro}}$-score metrics for classical, manual PQC and AQML-found PQC
		models. PQC$^{\mathrm{Solo}}_{\mathrm{AQML}}$ and PQC$^{\mathrm{Linear}}_{\mathrm{AQML}}$ denote
		PQC models without and with final classical linear layer respectively. We also provide trainable parameters
		number for each model.}
	\label{tab:mnist_results}
	\begin{tabular}{l|ccccc|c}
		\hline
		\textbf{Metric}       & \textbf{Fold 1} & \textbf{Fold 2} & \textbf{Fold 3} & \textbf{Fold 4} & \textbf{Fold 5} & \textbf{Mean}       \\
		\hline
		\multicolumn{7}{c}{MLP (\# Parameters. Feature extractors: 38372; Classifier: 37623, Total: 75995)}                                   \\
		\hline
		accuracy              & 0.962           & 0.959           & 0.960           & 0.960           & 0.960           & $0.960 \pm 0.001$   \\
		f1$_{\mathrm{macro}}$ & 0.961           & 0.959           & 0.959           & 0.959           & 0.959           & $0.959 \pm 0.001$   \\
		\hline
		\multicolumn{7}{c}{PQC$_{\mathrm{manual}}$ (\# Parameters. Feature extractors: 4614; Classifier: 18, Total: 4632)}                    \\
		\hline
		accuracy              & 0.880           & 0.855           & 0.880           & 0.905           & 0.911           & $0.886 \pm 0.022$   \\
		f1$_{\mathrm{macro}}$ & 0.879           & 0.852           & 0.879           & 0.903           & 0.910           & $0.885 \pm 0.023$   \\
		\hline
		\multicolumn{7}{c}{PQC$^{\mathrm{Solo}}_{\mathrm{AQML}}$ (\# Parameters. Feature extractors: 6202; Classifier: 72, Total: 6274)}      \\
		\hline
		accuracy              & 0.937           & 0.943           & 0.935           & 0.934           & 0.949           & $0.939 \pm 0.006$   \\
		f1$_{\mathrm{macro}}$ & 0.935           & 0.941           & 0.933           & 0.932           & 0.947           & $0.938 \pm 0.006$   \\
		\hline
		\multicolumn{7}{c}{PQC$^{\mathrm{Linear}}_{\mathrm{AQML}}$ (\# Parameters. Feature extractors: 10764; Classifier: 135, Total: 10899)} \\
		\hline
		accuracy              & 0.947           & 0.947           & 0.935           & 0.943           & 0.947           & $0.944 \pm 0.005$   \\
		f1$_{\mathrm{macro}}$ & 0.946           & 0.946           & 0.933           & 0.942           & 0.946           & $0.943 \pm 0.005$   \\
		\hline
	\end{tabular}
\end{table}

\subsection{ONERA}

\subsubsection{Dataset}
\begin{figure}
	\centering
	\includegraphics[scale=0.24]{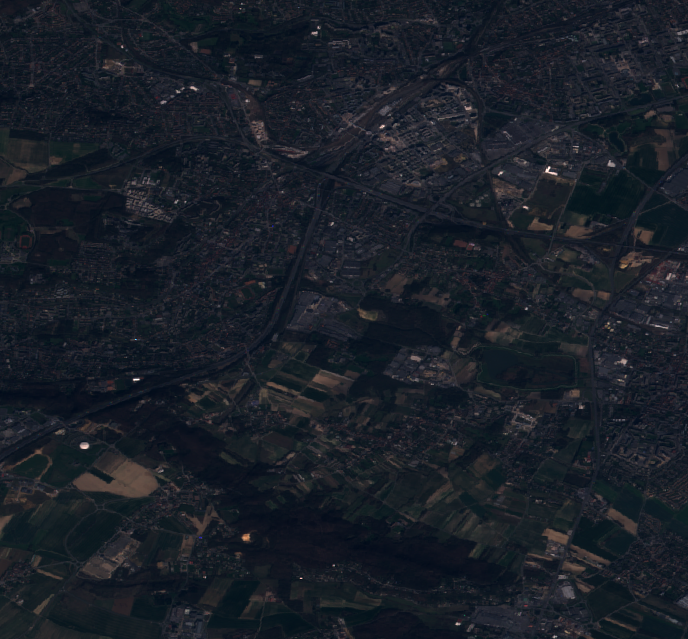}
	\includegraphics[scale=0.24]{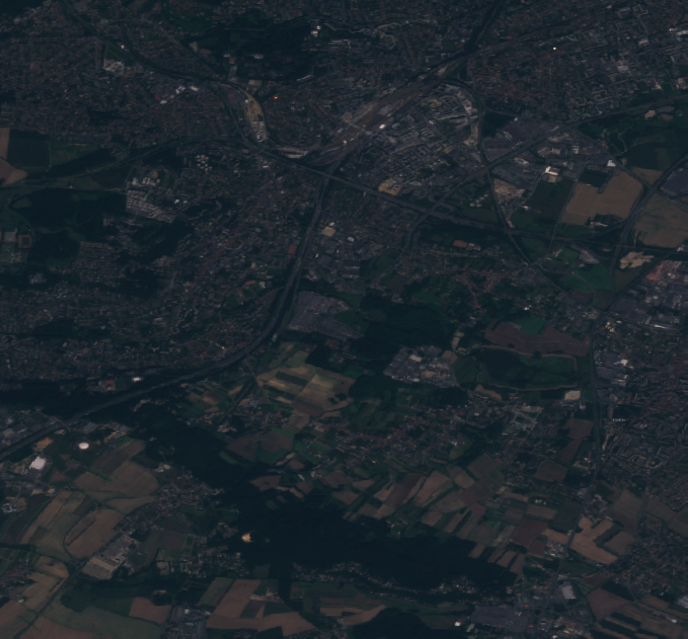}
	\caption{A pair of images from the ONERA dataset. They show Saclay, a city in France. The images were taken on 15
		III 2016 (Left) and 29 VIII 2017 (Right).}
	\label{fig:ONERA}
\end{figure}

The ONERA change detection dataset is a publicly available dataset containing annotated pairs of multispectral
satellite images \cite{onera}. The spatial resolution of the images vary between the pairs. Each image has 13 spectral
bands. An example of the input, reduced to the visible bands, is presented in the Figure~\ref{fig:ONERA}.

\subsubsection{Experimental setup}
In our experiments we repeat the experiment reported in \cite{QMLCaseStudy}, thus the setup remains the same. The only
change we include is using AQML to find ansatze used for the change detection. The metric we consider is still
accuracy and the data is preprocessed and fused in the same manner as in \cite{QMLCaseStudy}, as we explained in the
Section \ref{sec:method}.

For the AQML, we set up the number of QML blocks in range [1, 5]. We also included baseline classical classifier,
which was an MLP with 6 layers. As in \cite{QMLCaseStudy}, we didn't vary over the number of qubits, and it remained
equal to 8 for all the selected models.

\subsubsection{Results}

We ran 26, 36 and 41 experiments for MLP, PQC$^{\mathrm{Solo}}_{\mathrm{AQML}}$ and
PQC$^{\mathrm{Linear}}_{\mathrm{AQML}}$ models respectively. It's worth mentioning that in change detection
experiments the architecture of MLP didn't change from run to run. It is therefore justified to consider both the
base and the average performance of MLP in the comparison with PQC models.

While the accuracy of the best classical model outperformed the best PQC model --- 0.752 vs. 0.743 --- the best PQC
model was better than average MLP. The average performance comparison is even better, because the average accuracy
of PQC$^{\mathrm{Linear}}_{\mathrm{AQML}}$ models outperformed MLP, although only slightly  --- 0.730 vs. 0.728. Even
more surprising is the fact that the best PQC model found was a very basic model: \texttt{Amplitude} $\to$
$\texttt{BEL}$. This model had only 8 trainable parameters! For the comparison, the MLP we used had 384 trainable
parameters. While none of the models is close to the reported state-of-the-art, we conclude that given similar
restrictions, AQML-found PQCs lead to results comparable with classical models.

To further justify the comparison PQC and classical models, we rerun 13 repetitions of the experiment with MLP
containing a single hidden layer, reducing the number of trainable parameters to 64. The accuracy of MLP didn't
drop significantly, and remained at an impressive 0.731 on average (the best from all the models). However, the best
accuracy it achieved was 0.738, which is below the best accuracy reported by baseline MLP and PQC. Finishing this
direction of research, we considered an MLP with only 2 hidden neurons, and thus 16 parameters. Over another 10 runs,
the model scored 0.731, 0.667, 0.525 for maximal, average and minimal accuracy, all below AQML-found PQCs.

We also report that AQML approach led to the results that were better than originally reported (with accuracy
0.752 vs. 0.720). That's also true for PQC$^{\mathrm{Solo}}_{\mathrm{AQML}}$ models, which achieved best accuracy of
0.738. Furthermore, the best PQC model had only 8 parameters and significantly less gates than the model reported
in the original research. This conclusion is vital, because transpiling and running such model on a real device would
be much easier and would lead to lesser influence of the device imperfections. Models that simple were not even considered
in the original study! This highlights how useful the AQML approach can be, at least during the prototyping stage of the research.

Finally, we also report that while the additional classical linear layer had no significant influence on the best
accuracy obtained with both the models, we noticed that the stability of the training and the results are greatly
increased if we use the linear layer (and thus consider all the output qubits). The results are even more visible
than in the case of the MNIST dataset, with:
\begin{equation}
	\mathrm{accuracy}^{\mathrm{avg}}_{\mathrm{PQC}^{\mathrm{Solo}}_{AQML}}
	= 0.676 < 0.738 =
	\mathrm{accuracy}^{\mathrm{avg}}_{\mathrm{PQC}^{\mathrm{Linear}}_{AQML}},
\end{equation}
\begin{equation}
	\mathrm{accuracy}^{\mathrm{min}}_{\mathrm{PQC}^{\mathrm{Solo}}_{AQML}}
	= 0.505 < 0.710 =
	\mathrm{accuracy}^{\mathrm{min}}_{\mathrm{PQC}^{\mathrm{Linear}}_{AQML}}.
\end{equation}
This observation positions training stability research as an interesting direction for future investigations.

\section{Conclusion}
\label{sec:conclusions}

In this study, we provide evidence that even the simplest AQML approach can lead to better hybrid quantum-classical
models. This is particularly true for the models that also perform data fusion.

Using specially prepared MNIST as an synthetic example, we report that manually selected QML models were worse than
that found by AQML. Although the models were never as good as the best classical models, they were comparable, and
they needed far less trainable parameters to achieve their accuracy.

We also performed experiments using real data, this time using AQML to suggest the best QML model for the change detection
on ONERA dataset. Not only do we report finding a QML model better than the original work, but we also report, that on
average, the AQML-found models were better than classical MLPs with similar number of parameters. This result further
highlights the usefulness of the AQML approach.

An interesting by-product of our study is that the additional classical linear output layer seems to stabilize the
hybrid model training. We presented the preliminary results on that matter and leave the proper investigation for
further studies.

\vspace*{-0.3cm}
\begin{credits}

	\subsubsection{\ackname} We gratefully acknowledge the funding support by the program ``Excellence initiative—research
	university'' for the AGH University of Kraków as well as the ARTIQ project (UMO-2021/01/2/ST6/00004 and
	ARTIQ/0004/2021).

	\vspace*{-0.3cm}
	\subsubsection{\discintname}
	The authors have no competing interests to declare that are
	relevant to the content of this article.

\end{credits}

\setlength{\emergencystretch}{5em}
\bibliographystyle{splncs04}
\bibliography{aqml4msc.bib}

@software{dziura_2026_18717347,
	author       = {Dziura, Sebastian and
	Rybotycki, Tomasz and
	Gawron, Piotr},
	title        = {Automated Quantum Machine Learning for Multisource
	Classification: Experimental data and the code
	},
	month        = feb,
	year         = 2026,
	publisher    = {Zenodo},
	doi          = {10.5281/zenodo.18717347},
	url          = {https://doi.org/10.5281/zenodo.18717347},
}

@misc{ADAM,
	title={Adam: A Method for Stochastic Optimization}, 
	author={Diederik P. Kingma and Jimmy Ba},
	year={2017},
	eprint={1412.6980},
	archivePrefix={arXiv},
	primaryClass={cs.LG},
	url={https://arxiv.org/abs/1412.6980}, 
}

@misc{optuna,
	title={Optuna: A Next-generation Hyperparameter Optimization Framework}, 
	author={Takuya Akiba and Shotaro Sano and Toshihiko Yanase and Takeru Ohta and Masanori Koyama},
	year={2019},
	eprint={1907.10902},
	archivePrefix={arXiv},
	primaryClass={cs.LG},
	url={https://arxiv.org/abs/1907.10902}, 
}

@article{dalla_mura_challenges_2015,
	title = {Challenges and {Opportunities} of {Multimodality} and {Data} {Fusion} in {Remote} {Sensing}},
	volume = {103},
	issn = {1558-2256},
	url = {https://ieeexplore.ieee.org/abstract/document/7194740},
	doi = {10.1109/JPROC.2015.2462751},
	number = {9},
	urldate = {2026-02-19},
	journal = {Proceedings of the IEEE},
	author = {Dalla Mura, Mauro and Prasad, Saurabh and Pacifici, Fabio and Gamba, Paulo and Chanussot, Jocelyn and Benediktsson, Jón Atli},
	month = sep,
	year = {2015},
	pages = {1585--1601},
}

@misc{schuld_effect_2021,
	title = {The effect of data encoding on the expressive power of variational quantum machine learning models},
	doi = {10.1103/PhysRevA.103.032430},
	urldate = {2026-02-19},
	author = {Schuld, Maria and Sweke, Ryan and Meyer, Johannes Jakob},
	month = mar,
	year = {2021},
	note = {arXiv:2008.08605 [quant-ph]},
}

@article{hancock_survey_2020,
	title = {Survey on categorical data for neural networks},
	volume = {7},
	issn = {2196-1115},
	doi = {10.1186/s40537-020-00305-w},
	language = {en},
	number = {1},
	urldate = {2026-02-18},
	journal = {Journal of Big Data},
	author = {Hancock, John T. and Khoshgoftaar, Taghi M.},
	pages = {28},
}

@book{RussellNorvig2010,
	author    = {Stuart Russell and Peter Norvig},
	title     = {Artificial Intelligence: A Modern Approach},
	publisher = {Prentice Hall},
	year      = {2010},
	edition   = {3rd},
	isbn      = {978-0136042594}
}

@INPROCEEDINGS{onera,
	author={Caye Daudt, Rodrigo and Le Saux, Bertr and Boulch, Alexandre},
	booktitle={2018 25th IEEE International Conference on Image Processing (ICIP)}, 
	title={Fully Convolutional Siamese Networks for Change Detection}, 
	year={2018},
	volume={},
	number={},
	pages={4063-4067},
	doi={10.1109/ICIP.2018.8451652}
}

@misc{ChangeDetectionSurvey,
	title={Change Detection Methods for Remote Sensing in the Last Decade: A Comprehensive Review}, 
	author={Guangliang Cheng and Yunmeng Huang and Xiangtai Li and Shuchang Lyu and Zhaoyang Xu and Qi Zhao and Shiming Xiang},
	year={2023},
	eprint={2305.05813},
	archivePrefix={arXiv},
	primaryClass={cs.CV},
	url={https://arxiv.org/abs/2305.05813}, 
}

@misc{Segmentation,
	title={Semantic Image Segmentation: Two Decades of Research}, 
	author={Gabriela Csurka and Riccardo Volpi and Boris Chidlovskii},
	year={2023},
	eprint={2302.06378},
	archivePrefix={arXiv},
	primaryClass={cs.CV},
	url={https://arxiv.org/abs/2302.06378}, 
}

@misc{QMLCaseStudy,
	title={Explainable Quantum Machine Learning for Multispectral Images Segmentation: Case Study}, 
	author={Tomasz Rybotycki and Manish K. Gupta and Piotr Gawron},
	year={2025},
	eprint={2503.08962},
	archivePrefix={arXiv},
	primaryClass={quant-ph},
	url={https://arxiv.org/abs/2503.08962}, 
}

@inproceedings{QAS,
	title={Quantum Architecture Search: A Survey},
	DOI={10.1109/qce60285.2024.00198},
	booktitle={2024 IEEE International Conference on Quantum Computing and Engineering (QCE)},
	publisher={IEEE},
	author={Martyniuk, Darya and Jung, Johannes and Paschke, Adrian},
	year={2024},
	month=sep, pages={1695–1706} }

@article{Rybotycki_2025,
	title={{AQMLator – an Quantum Machine Learning e-Platform}},
	volume={26},
	ISSN={1508-2806},
	DOI={10.7494/csci.2025.26.si.7063},
	number={SI},
	journal={Computer Science},
	publisher={AGH University Press},
	author={Rybotycki, Tomasz and Gawron, Piotr},
	year={2025},
	month=jul }

@misc{AQML_Exp,
	title={AutoQML: Automated Quantum Machine Learning for Wi-Fi Integrated Sensing and Communications}, 
	author={Toshiaki Koike-Akino and Pu Wang and Ye Wang},
	year={2022},
	eprint={2205.09115},
	archivePrefix={arXiv},
	primaryClass={cs.LG},
	url={https://arxiv.org/abs/2205.09115}, 
}

@ARTICLE{GawronPowerOfData,
	author={Gupta, Manish K. and Romaszewski, Michał and Gawron, Piotr},
	volume={73},
	number={5},
	journal={Bulletin of the Polish Academy of Sciences Technical Sciences},
	pages={e154279},
	howpublished={online},
	year={2025},
	type={article},
	title={Potential of quantum machine learning for processing multispectral Earth observation data},
	doi={10.24425/bpasts.2025.154279},
}

@ARTICLE{DataFusionNPQAOA,
	author={Stooβ, Veit and Ulmke, Martin and Govaers, Felix},
	journal={IEEE Transactions on Aerospace and Electronic Systems}, 
	title={Quantum Computing for Applications in Data Fusion}, 
	year={2023},
	volume={59},
	number={2},
	pages={2002-2012},
	doi={10.1109/TAES.2022.3212026}}

@ARTICLE{DataFusion1986,
	author={Chair, Z. and Varshney, P.K.},
	journal={IEEE Transactions on Aerospace and Electronic Systems}, 
	title={Optimal Data Fusion in Multiple Sensor Detection Systems}, 
	year={1986},
	volume={AES-22},
	number={1},
	pages={98-101},
	doi={10.1109/TAES.1986.310699}}

@article{abbasPowerQuantumNeural2021,
  title     = {The Power of Quantum Neural Networks},
  author    = {Abbas, Amira and Sutter, David and Zoufal, Christa and Lucchi, Aurelien and Figalli, Alessio and Woerner, Stefan},
  year      = 2021,
  month     = jun,
  journal   = {Nature Computational Science},
  volume    = {1},
  number    = {6},
  pages     = {403--409},
  publisher = {Nature Publishing Group},
  issn      = {2662-8457},
  doi       = {10.1038/s43588-021-00084-1},
  urldate   = {2026-02-05},
  copyright = {2021 The Author(s), under exclusive licence to Springer Nature America, Inc.},
  langid    = {english}
}

@article{baltrusaitisMultimodalMachineLearning2019,
  title      = {Multimodal {{Machine Learning}}: {{A Survey}} and {{Taxonomy}}},
  shorttitle = {Multimodal {{Machine Learning}}},
  author     = {Baltru{\v s}aitis, Tadas and Ahuja, Chaitanya and Morency, Louis-Philippe},
  year       = 2019,
  month      = feb,
  journal    = {IEEE Transactions on Pattern Analysis and Machine Intelligence},
  volume     = {41},
  number     = {2},
  pages      = {423--443},
  issn       = {1939-3539},
  doi        = {10.1109/TPAMI.2018.2798607},
  urldate    = {2026-01-29}
}

@article{biamonteQuantumMachineLearning2017a,
  title     = {Quantum Machine Learning},
  author    = {Biamonte, Jacob and Wittek, Peter and Pancotti, Nicola and Rebentrost, Patrick and Wiebe, Nathan and Lloyd, Seth},
  year      = 2017,
  month     = sep,
  journal   = {Nature},
  volume    = {549},
  number    = {7671},
  pages     = {195--202},
  publisher = {Nature Publishing Group},
  issn      = {1476-4687},
  doi       = {10.1038/nature23474},
  urldate   = {2026-02-03},
  copyright = {2017 Macmillan Publishers Limited, part of Springer Nature. All rights reserved.},
  langid    = {english}
}

@article{fortinoAdvancesMultisensorFusion2019,
  title      = {Advances in Multi-Sensor Fusion for Body Sensor Networks: {{Algorithms}}, Architectures, and Applications},
  shorttitle = {Advances in Multi-Sensor Fusion for Body Sensor Networks},
  author     = {Fortino, Giancarlo and Ghasemzadeh, Hassan and Gravina, Raffaele and Liu, Peter X. and Poon, Carmen C. Y. and Wang, Zhelong},
  year       = 2019,
  month      = jan,
  journal    = {Information Fusion},
  volume     = {45},
  pages      = {150--152},
  issn       = {1566-2535},
  doi        = {10.1016/j.inffus.2018.01.012},
  urldate    = {2026-02-02}
}

@article{havlicekSupervisedLearningQuantumenhanced2019,
  title   = {Supervised Learning with Quantum-Enhanced Feature Spaces},
  author  = {Havl{\'i}{\v c}ek, Vojt{\v e}ch and C{\'o}rcoles, Antonio D. and Temme, Kristan and Harrow, Aram W. and Kandala, Abhinav and Chow, Jerry M. and Gambetta, Jay M.},
  year    = 2019,
  month   = mar,
  journal = {Nature},
  volume  = {567},
  number  = {7747},
  pages   = {209--212},
  issn    = {0028-0836, 1476-4687},
  doi     = {10.1038/s41586-019-0980-2},
  urldate = {2025-12-15},
  langid  = {english}
}

@article{huangPowerDataQuantum2021a,
  title     = {Power of Data in Quantum Machine Learning},
  author    = {Huang, Hsin-Yuan and Broughton, Michael and Mohseni, Masoud and Babbush, Ryan and Boixo, Sergio and Neven, Hartmut and McClean, Jarrod R.},
  year      = 2021,
  month     = may,
  journal   = {Nature Communications},
  volume    = {12},
  number    = {1},
  pages     = {2631},
  publisher = {Nature Publishing Group},
  issn      = {2041-1723},
  doi       = {10.1038/s41467-021-22539-9},
  urldate   = {2026-02-05},
  copyright = {2021 The Author(s)},
  langid    = {english}
}

@misc{lecunMNISTHandwrittenDigits2025,
  title   = {{{MNIST Handwritten Digits}}},
  author  = {LeCun, Yann},
  year    = 2025,
  month   = dec,
  url     = {https://www.kaggle.com/datasets/hichamachahboun/mnist-handwritten-digits},
  address = {https://www.kaggle.com/datasets/hichamachahboun/mnist-handwritten-digits}
}

@article{liA3CLNNSpatialSpectral2022,
  title         = {{{A3CLNN}}: {{Spatial}}, {{Spectral}} and {{Multiscale Attention ConvLSTM Neural Network}} for {{Multisource Remote Sensing Data Classification}}},
  shorttitle    = {{{A3CLNN}}},
  author        = {Li, Heng-Chao and Hu, Wen-Shuai and Li, Wei and Li, Jun and Du, Qian and Plaza, Antonio},
  year          = 2022,
  month         = feb,
  journal       = {IEEE Transactions on Neural Networks and Learning Systems},
  volume        = {33},
  number        = {2},
  eprint        = {2204.04462},
  primaryclass  = {cs},
  pages         = {747--761},
  issn          = {2162-237X, 2162-2388},
  doi           = {10.1109/TNNLS.2020.3028945},
  urldate       = {2026-02-02},
  archiveprefix = {arXiv}
}

@misc{sebastianelli2021circuitbasedhybridquantumneural,
	title={On Circuit-based Hybrid Quantum Neural Networks for Remote Sensing Imagery Classification}, 
	author={Alessandro Sebastianelli and Daniela A. Zaidenberg and Dario Spiller and Bertrand Le Saux and Silvia Liberata Ullo},
	year={2021},
	eprint={2109.09484},
	archivePrefix={arXiv},
	primaryClass={eess.IV},
	url={https://arxiv.org/abs/2109.09484}, 
}

@article{liMultisourceInformationFusion2024,
  title      = {Multi-Source Information Fusion: {{Progress}} and Future},
  shorttitle = {Multi-Source Information Fusion},
  author     = {Li, Xinde and Dunkin, Fir and Dezert, Jean},
  year       = 2024,
  month      = jul,
  journal    = {Chinese Journal of Aeronautics},
  volume     = {37},
  number     = {7},
  pages      = {24--58},
  issn       = {10009361},
  doi        = {10.1016/j.cja.2023.12.009},
  urldate    = {2024-12-18},
  langid     = {english}
}

@article{schuldQuantumMachineLearning2019,
  title         = {Quantum Machine Learning in Feature {{Hilbert}} Spaces},
  author        = {Schuld, Maria and Killoran, Nathan},
  year          = 2019,
  month         = feb,
  journal       = {Physical Review Letters},
  volume        = {122},
  number        = {4},
  eprint        = {1803.07128},
  primaryclass  = {quant-ph},
  pages         = {040504},
  issn          = {0031-9007, 1079-7114},
  doi           = {10.1103/PhysRevLett.122.040504},
  urldate       = {2024-12-23},
  archiveprefix = {arXiv}
}

@article{sycaraIntegratedApproachHighlevel2009,
  title     = {An Integrated Approach to High-Level Information Fusion},
  author    = {Sycara, Katia and Glinton, Robin and Yu, Bin and Giampapa, Joseph and Owens, Sean and Lewis, Michael and Grindle, Ltc Charles},
  year      = 2009,
  month     = jan,
  journal   = {Information Fusion},
  volume    = {10},
  number    = {1},
  pages     = {25--50},
  issn      = {15662535},
  doi       = {10.1016/j.inffus.2007.04.001},
  urldate   = {2026-02-02},
  copyright = {https://www.elsevier.com/tdm/userlicense/1.0/},
  langid    = {english}
}

@misc{wuFeatureEntanglementbasedQuantum2026,
  title         = {Feature {{Entanglement-based Quantum Multimodal Fusion Neural Network}}},
  author        = {Wu, Yu and Zhou, Qianli and Geng, Jie and Deng, Xinyang and Jiang, Wen},
  year          = 2026,
  month         = jan,
  number        = {arXiv:2601.07856},
  eprint        = {2601.07856},
  primaryclass  = {quant-ph},
  publisher     = {arXiv},
  doi           = {10.48550/arXiv.2601.07856},
  urldate       = {2026-02-02},
  archiveprefix = {arXiv}
}

\end{document}